# An alternative ultrasonic method for measuring the elastic properties of cortical bone

Pithioux M., Lasaygues P., Chabrand P.

Laboratoire de Mécanique et d'Acoustique, CNRS, 31 Ch. Joseph Aiguier 13402 Marseille France.
mail : pithioux@lma.cnrs-mrs.fr
tel : +33 (0)4 91 16 44 38
tel : +33 (0)4 91 16 44 81

# An alternative ultrasonic method for measuring the elastic properties of cortical bone

Pithioux M., Lasaygues P., Chabrand P.
Laboratoire de Mécanique et d'Acoustique, CNRS, 31 Ch. Joseph Aiguier 13402 Marseille France.
mail: pithioux@lma.cnrs-mrs.fr

ABSTRACT

We studied the elastic properties of bone to analyze its mechanical behavior. The basic principles of ultrasonic methods are now well established for varying isotropic media, particularly in the field of biomedical engineering. However, little progress has been made in its application to anisotropic materials. This is largely due to the complex nature of wave propagation in these media. In the present study, the theory of elastic waves is essential because it relates the elastic moduli of a material to the velocity of propagation of these waves along arbitrary directions in a solid. Transducers are generally placed in contact with the samples which are often cubes with parallel faces that are difficult to prepare. The ultrasonic method used here is original, a rough preparation of the bone is sufficient and the sample is in rotation. Moreover, to analyze heterogeneity of the structure we measure velocities in different points on the sample. The aim of the present study was to determine in vitro the anisotropic elastic properties of cortical bones. For this purpose, our method allowed measuring longitudinal and transversal velocities ($C_L$ and $C_T$) in longitudinal (fiber direction) and radial directions (orthogonal to the fiber direction) of compact bones. Young's modulus E and Poisson's ratio ν, were then deduced from the velocities measured considering the compact bone as transversely isotropic or orthotropic. The results are in line with those of other methods.

INTRODUCTION

The present study develops a new experimental ultrasonic procedure for measuring in vitro the elastic characteristics of compact bones. Methods of this kind have been used to study compact bones (Ashman et al., 1984; Bonfield and Tully, 1982; Fung, 1981; Hobatho et al., 1998; Lowet and Van der Perre, 1996; Mehta, et al., 1999; Sedlin, 1966; Viano, 1995). The ultrasonic method used here is original; the previous techniques used transducers in contact with the samples; and samples are often cubes with parallel faces that are difficult to prepare. In this case, it is difficult to cut cubes parallel to the fibers which are oriented at different angles in comparison with the principal axis of the bone. In our case, a rough preparation of the bone is sufficient, the form of the samples is that of the bones and the sample is in rotation. Herein, waves propagating both longitudinally and transversally in all the plane directions of bovine femoral bones were used to characterize their mechanical properties. The applications of the ultrasonic control system first developed at the Laboratoire de Mécanique et d'Acoustique (LMA) for ultrasonic medical imaging (Lefebvre, 1988; Lasaygues and Lefebvre, 2001) were extended to include the elastic properties of biological systems. To determine the elastic properties of a solid, the theory of elastic waves is essential because it relates the elastic moduli of a material to the speed of propagation of these waves along arbitrary directions in a solid. The aim of this work is to describe the inversion of ultrasonic wavespeed measurements on a bovine bone to represent its elastic properties using a rapid and precise method.

This experimental method based on theoretical works (Castagnede et al., 1990; Sachse et al., 1998), was adapted to the problem of a transversely isotropic material such as unidirectional fiber-reinforced composites (Lasaygues, 1992) and more recently to the problem of bone quantitative imaging (Lasaygues et al., 2001).

The bench was designed for performing both reflection and transmission measurements. Our mechanical acoustic device allows various degrees of freedom, since the position of both the target and the transducers can be adjusted. In particular, one can prescribe rotation on the target, and the transducer receiver can be moved laterally. This makes it possible to monitor the wave propagation in a system that obeys Snell-Descartes laws. The mechanical parameters of each specimen were determined at various points on specimen bones to determine the heterogeneity of the samples.

1. THEORETICAL APPROACH: IDENTIFICATION OF ELASTIC CONSTANTS

In this study, we focused on bovine bones, considered an elastic medium. Bovine bones have a lamellar structure and are generally considered orthotropic. However, we propose in the future to investigate human bones, which are mainly Haversian, and which are usually considered to be transversely isotropic (Ashman et al., 1984; Yoon and Katz, 1976). We will, therefore, investigate bovine bone using theory appropriate for both symmetries.

1.1 Elastic constants considering the bone as transversely isotropic or orthotropic.

For generally anisotropic media, Hooke law is written:

$$\sigma_{ij}(x) = C_{ijkl}(x)\varepsilon_{kl}(x) \quad \text{where} \quad (i,j,k,l \in \{1,2,3\}) \tag{1}$$

In (1) $\sigma_{ij}$ denotes the ij component of the stress tensor, and $\varepsilon_{kl}$ represents the components of the strain infinitesimal tensor. The 21 coefficients $C_{ijkl}$ in (1) characterize the stiffness matrix elastic constants. Only five or nine independent elements were determined, however, because cortical bone microstructure has a hexagonal symmetry.

1.1.1 Material considered as transversely isotropic

Axis $(0X_3)$ was taken to be the fiber axis. The material was assumed isotropic in the $(0X_1X_2)$ plane (perpendicular to the fibers). By determining the propagation in a direction in the plane $(0X_1X_2)$, we can therefore measure the longitudinal and transversal velocities in this plane and deduce constants $C_{11}$, $C_{66}$, (Rose, 1999) such as:

$$C_{11} = \rho C_L^2$$

$$C_{66} = \rho C_T^2$$

$C_{12}$ is then given by $C_{12} = C_{11} - 2C_{66}$

In the plane $(0X_1X_3)$, which contains fibers, and for any direction $\vec{p} = (\cos r, 0, \sin r)$, let D be the symmetric matrix whose coefficients are given by:

$$\begin{cases} D_{11} = C_{11} \cos^2 r + C_{44} \sin^2 r - \lambda \\ D_{12} = D_{23} = 0 \\ D_{13} = (C_{13} + C_{44}) \cos r \sin r \\ D_{22} = C_{66} \cos^2 r + C_{44} \sin^2 r - \lambda \\ D_{33} = C_{44} \cos^2 r + C_{33} \sin^2 r - \lambda \end{cases} \quad (2)$$

In (2), r is the known refraction angle of the transmitted wave. The eigenvalues of D are given by $\lambda = \rho C^2$. We then make the following change of variables: $X_1 = C_{33}$; $X_2 = C_{44}$; $X_3 = C_{13}$; $a = C_{11}$

To measure k, $k \in \{1,..,N\}$, with N the total number of measurements, angle $r_k$, longitudinal or transversal velocity $C_k$ we write: $\alpha_k = \cos^2 r_k$ and $b_k = \rho C_k^2$

And after making some calculations, we have:

$$\det(D_k) = f_k(X_1, X_2, X_3) = A_1 X_1 + A_2 X_2 + A_3 X_3^2 + A_4 X_1 X_2 + 2 A_3 X_2 X_3 + A_5 = 0 \qquad (3)$$

with 
$$\begin{cases} A_1 = (1 - \alpha_k)(a\alpha_k - b_k) \\ A_2 = a\alpha_k^2 - b_k \\ A_3 = -\alpha_k (1 - \alpha_k) \\ A_4 = (1 - \alpha_k)^2 \\ A_5 = -b_k (a\alpha_k - b_k) \end{cases}$$

Equation (3) is a system of N equations with three unknowns. To solve the problem, we choose an Euclidean norm and the functional:

$$F(X) = \sum_k \beta_k (f_k(X))^2 \text{ where } 0 \leq \beta_k \leq 1$$

$\beta_k$ is a weighting factor for increasing or decreasing the influence of k. X is the vector of the unknown, minimized using a Newton method.

1.1.2 Material considered as orthotropic:

When the compact bone is considered as orthotropic, the plane $(0X_1 X_3)$, contains fibers, for any direction $\vec{p} = (\cos r, 0, \sin r)$. We have D, which becomes:

$$\begin{cases} D_{11} = C_{11} \cos^2 r + C_{55} \sin^2 r - \lambda \\ D_{12} = D_{23} = 0 \\ D_{13} = (C_{13} + C_{55}) \cos r \sin r \\ D_{22} = C_{66} \cos^2 r + C_{44} \sin^2 r - \lambda \\ D_{33} = C_{55} \cos^2 r + C_{33} \sin^2 r - \lambda \end{cases} \qquad (4)$$

With $\det(D) = 0$. We then make the following change of variables $X_1 = C_{33}$; $X_2 = C_{55}$; $X_3 = C_{13}$; $a = C_{11}$

Once these coefficients are determined, we can deduce the others. If we determine the propagation in a direction ($0X_1$) (resp ($0X_2$)), we can measure longitudinal and transversal velocities and deduce constants $C_{11}$, $C_{66}$, (resp $C_{22}$, $C_{44}$) such as :

$$C_{11} = \rho C_L^2 \text{ in } (0X_1) \qquad (5)$$

$$C_{66} = \rho C_T^2 \text{ in } (0X_1) \qquad (6)$$

$$C_{22} = \rho C_L^2 \text{ in } (0X_2) \qquad (7)$$

$$C_{44} = \rho C_T^2 \text{ in } (0X_2) \qquad (8)$$

Finally, coefficients $C_{12}$ (resp $C_{23}$) were deduced when the samples were at an angle of 45° in the plane ($0X_1X_2$) (resp ($0X_2X_3$)).

The $X_i$ determined are good approximations of the zeros of the functional F. From all the calculations, we obtain $\max_{i \leq 3} |F(X_i)| \leq 10^{-11}$.

## 2. MATERIALS - METHODS

2.1 Experiments

It is clear from the theory presented before that wavespeed data should be collected over a broad range of arbitrary directions in a specimen in order to recover the whole set of elastic constants.

The general architecture (Fig. 1) of the mechanical system is composed of a main symmetric arm holding two transverse arms that move two transducers in parallel. Angular scanning is carried out by rotating either the main arm or the object holder. The transducers can also be positioned and oriented precisely; such precision allows for both linear and sectorial scanning. All the movements are produced by six stepping motors sequentially driven by a programmable translator-indexer device fitted with a power multiplexer. The translator-indexer and the power multiplexer are integrated in a control rack with other remote controls, such as that for adjusting the distance traveled by the transverse arms, or for the out-of-water setting. The increments are multiples of 0.75 $10^{-2}$ millimeters for translations and of 1 $10^{-2}$ degrees for rotations (Lasaygues and Lefebvre, 2001). 3 mm focused transducers were used with a central frequency of 1 MHz.

Ten fresh bovine femoral bones were studied. The bones were frozen prior to the experiments. The epiphyses were cut off so that we could concentrate our attention only on compact bone. The method does not require the use of samples with precise dimensions and perfectly parallel faces: rough preparations of the specimen are sufficient. The area under investigation must have interfaces, which are approximately parallel (focus of the transducers).

Two test-pieces were obtained by first cutting bones in the axial direction and then removing the marrow from each part (Fig. 5.a-e). The sample, set in water at room temperature, was held by the robot in either the horizontal or vertical position, depending on the type of experiment. Four series of measurements were performed on each sample: two reflection series (one on each longitudinal side) to determine the acoustical thickness, and two transmission series (one with the target motionless and one with it rotating) to determine the longitudinal and transversal velocity.

2.2 Acoustic measurement of the bone thickness.

The bone thickness was first calculated using the echo technique with transducers used as both the transmitter and receiver. Let d be the distance between the two transducers, $t_1$ and $t_2$ be the time taken by the reflected echo to travel between each of the bone surfaces, and $C_w$ be the water velocity (Fig. 2.a).

The acoustical thickness x is then given by the following relation:

$$x = d - C_w \left( \frac{t_1 + t_2}{2} \right) \quad (9)$$

2.3. Calculating the longitudinal and transversal velocity

We investigated the transmission mode to determine the longitudinal and transversal velocities. For this purpose, pure compression waves were emitted. The longitudinal and transversal waves in the test-piece were then determined versus the incidence angle i. First, a reference measurement was obtained without any bone sample to determine $C_w$, $t_w$. Then with the bone sample the longitudinal and transversal velocities were determined by rotating the bone around the clamp axis so as to change the incidence of the acoustic beam. The emitter was fixed and the receiver could be moved laterally (Fig. 2.b). We define $i_c$ as the critical angle such that:

$$i_c = \arcsin \frac{C_w}{C_L} \quad (10)$$

When $i < i_c$, we can observe the longitudinal waves and when $i > i_c$, we can see the transversal ones.

The analyzed zone corresponds to the lateral resolution of the focused transducers (3 mm) and we assume that locally the input-output interfaces of the wave are parallel (Fig. 3). We can therefore apply Snell-Descartes laws.

According to the Snell Descartes laws, the refraction index is given by $n = \frac{C_w}{C} = \frac{\sin i}{\sin r}$ (11)

where i is the angle of incidence of the emitted wave and r the refraction angle of the transmitted wave.

And $\Delta t = \frac{x}{C_w} \left( \cos i - \sqrt{n^2 - \sin^2 i} \right)$ (12)

where x is the sample thickness. For every angle, Δt is obtained by cross correlation between the reference signal without the sample and the signal obtained at the angle of incidence with the sample (Fig. 4). The image was constructed with all transmission signals through a sample. The amplitude of each signal is grey scale coding.

From (11) and (12), $$C = \frac{C_w}{\sqrt{1 + \frac{C_w \Delta t}{x}\left(\frac{C_w \Delta t}{x} - 2\cos i\right)}} \qquad (13)$$

where C is the longitudinal or transversal bone velocity, depending on the experimental situation; $C_L$ and $C_T$ are therefore determined for several angles i.

## 3. RESULTS

The results are presented considering compact bone as transversely isotropic or orthotropic.

### 3.1. Transversal and longitudinal velocities in two bone directions (transversely isotropic structure).

To obtain accurate results by taking bone heterogeneity into account, we used a 1 MHz frequency with focused transducers. Focused transducers made it possible to focus on small surfaces and to perform measurements at different points (Fig. 5) on the bone sample. In this case, $C_T$ was measured in the two radial directions of the bone. To be able to observe transversal and longitudinal waves separately and because the critical angle $i_c$ was between $25° \leq i_c \leq 40°$, we prescribed series of one-degree rotation on the bone samples with a total angular rotation of 60°. First, to accurately describe the bone heterogeneity, we took ultrasonic measurements in various points, brought closer to each other (Fig. 5), so the heterogeneity of the medium could be taken rigorously into account. Two examples of results are presented with samples in the vertical and horizontal positions (Table 1). In parallel to this, we measured $C_L$ and $C_T$ of the two samples positioned in radial axes (Table 1).

Considering the $C_T$, compact zone of the bovine bones was weakly heterogeneous (for example, we obtained the same values for three points of the bone 2 in horizontal position).

In another experimental setup, we measured the velocities $C_L$ and $C_T$ in radial axes and in the longitudinal axis of two other fresh bovine bones (bone 3 and 4) (Table 2). The transversely isotropic hypothesis was adopted in this case. The results are summarized in the first column of the tables 2 and 3.

3.2. Transversal and longitudinal velocities in three bone directions (orthotropic structure).

Here both $C_L$ and $C_T$ were measured in the two radial and the longitudinal axes of the bone. In this case, the orthotropic hypothesis was adopted. Measurements were made in bones 3 and 4. The results are summarized in the second column of the tables 2 and 3. The matrix rigidity values in 3.1 and 3.2 were found to be similar and reproducible. The radial values showed low variability, as those in 3.1. When we considered the compact bone as transversely isotropic or orthotropic, the longitudinal velocities in the direction of the fibers (between 4000 m/s and 4400 m/s) were greater than those in the radial direction (between 3000 m/s and 3600 m/s). On the contrary, we had a similar variation of the transversal velocities in the radial direction (between 1700 m/s and 2100 m/s) and in the direction of the fibres (between 1900 m/s and 2100 m/s).

4. DISCUSSION AND CONCLUSION

In the literature (Ashman, et al., 1984; Bonfield and Tully, 1982; Fung, 1981; Katz et al., 1979, 1984; Lipson et al., 1984; Reilly and Burstein, 1975; Viano, 1995, Yoon and Katz, 1976), the coefficients of matrix rigidity and longitudinal velocity values for compact bone vary widely. The longitudinal velocity values have varied between 2700 m/s and 4200 m/s, ours ranged between 2900 m/s and 4400 m/s. The elastic constants were also compared with those in the literature (Table 4).

In the experiments, only a few measurement points were used so it was difficult to finely analyze the distribution of the bone characteristics in different regions of the samples. To overcome this problem, one approach consists in making a map of longitudinal and transversal velocities of the sample.

We did not study the viscosity of the sample. To determine whether any dispersion occurs in bones, such a study should determine the velocities and attenuation at various frequencies.

Until now, the elastic characteristics of bones have been measured by mechanical tests (compression, tension, flexion, torsion). With these tests, there is a risk of damaging the sample, rendering it unsuitable for tests in other directions or for other measurements. In this study, the elastic characteristics of compact bone were measured using an ultrasonic method. One of the advantages of the present method is that it gives fast accurate results. Moreover it does not require that samples have precise dimensions and perfectly parallel faces. Rough preparations of the specimen are sufficient. The only requirement is for the area under investigation to have interfaces that are approximately parallel. We consistently used the same experimental procedure on bone samples from animals of the same age (about five years), sex (female), and weight (about 4 hundred kilos). The ultrasonic method proposed here is an original approach to the study of bone characteristics because the bone is free to rotate around the clamp axis. This makes it possible to monitor the wave propagation, to measure shear waves, and to determine the velocities of these waves ($C_T$) in all the longitudinal and radial directions and for all the angles of rotation of the samples. The latter point is an original feature that classical ultrasonic methods do not provide.

The longitudinal velocity and coefficients of matrix rigidity obtained here are in line with those in the literature.

We also intend to apply these results to construct quantitative images of compact, cancellous bones and osteoporosis bones (Lasaygues and Lefebvre, 2001). Lastly, the results of this study may be used in a numerical model of bones that was developed to analyze failure in this structure (Pithioux, 2000).

List of captions

Fig. 1: Schematic diagram of the ultrasonic scanner. The relative displacements between the probe and the object were applied along the X, Y, and Z axes.

Fig. 2: Principle of the wave recording in the echo (a) and transmission modes (b)

Fig. 3: (a) Qualitative ultrasonic image of the sample. Horizontal section at the height z.

(b) Shape of the samples studied.

Fig. 4: Set of transmission signals versus incidence angle obtained with one example of sample at 1MHz. The image was constructed with all transmission signals through a sample. The amplitude of each signal is grey scale coding.

Fig. 5: Samples (a: Sample in vertical position; b: Sample in horizontal position; c: Study in the radial direction (vertical position); d: Study in the radial direction (horizontal position); e: Study in the longitudinal direction (axis 0x)). Measurements were carried out at several points called A, B, 1-6.

Table 1: Transversal and longitudinal measurements of velocities traveling in radial bone directions. A logarithmic differential of equation (13) was apply to calculate the error.

Table 2: Transversal and longitudinal measurements considering compact bone as transversely isotropic or orthotropic. . A logarithmic differential of equation (13) was apply to calculate the error.

Table 3: Elastic constants considering the compact bone as transversely isotropic or orthotropic.

Table 4: Comparison with data in the literature on the elastic constants. The results taken in the literature are all obtained with the same species (bovine bones), but the age, sex, and weight of the bovine bones were not given in all the papers.

List of illustrations

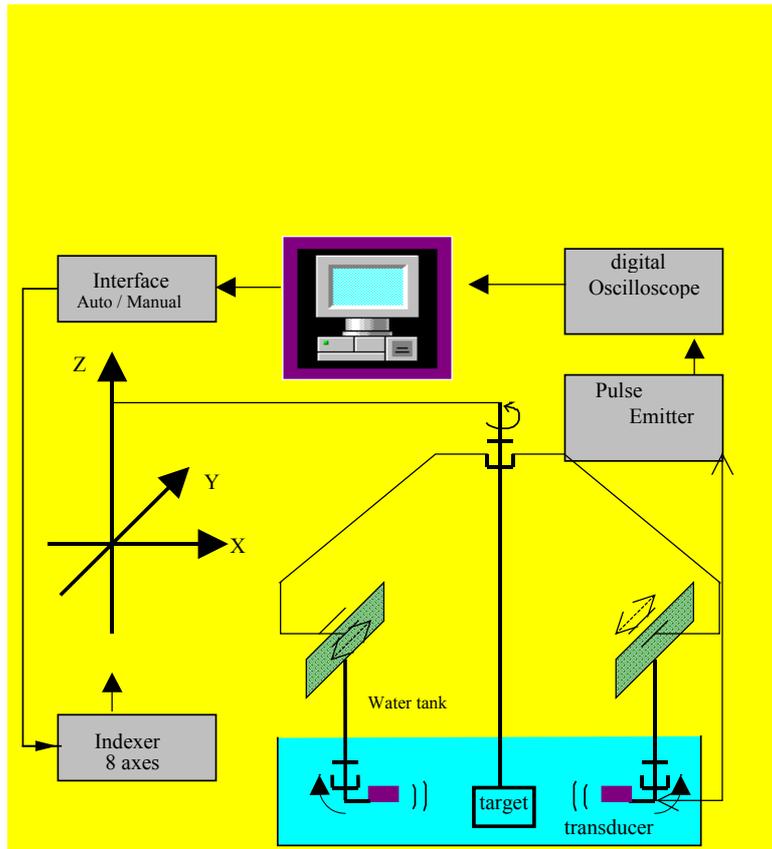

Fig.1

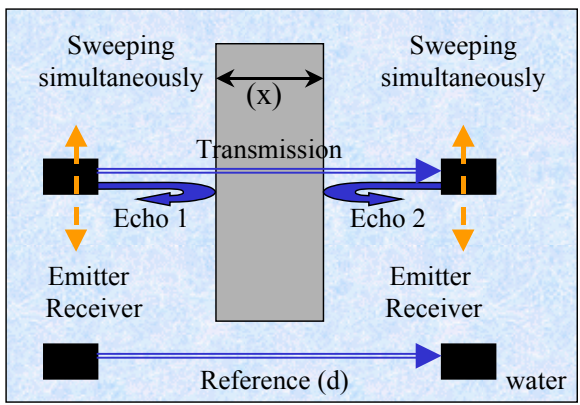
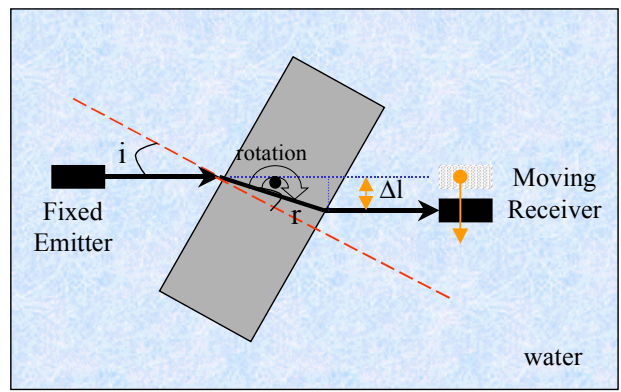

(a): Echo mode            (b): Transmission mode

Fig.2

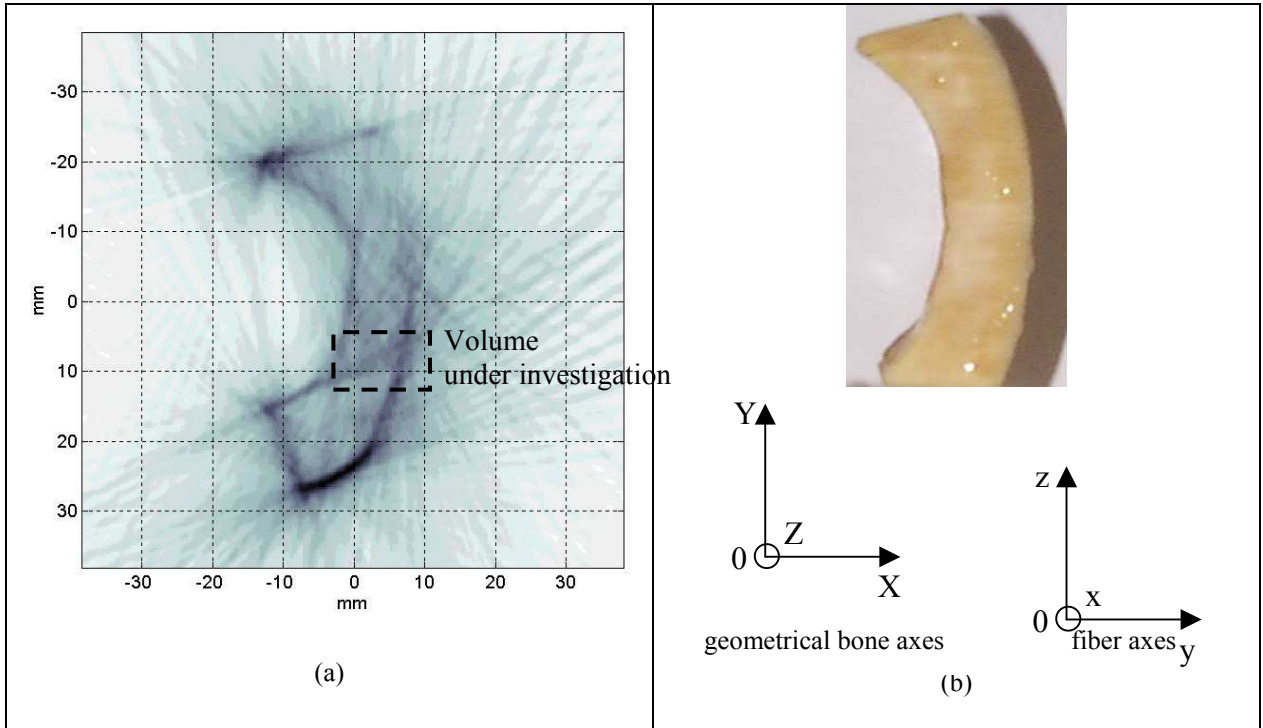

Fig.3

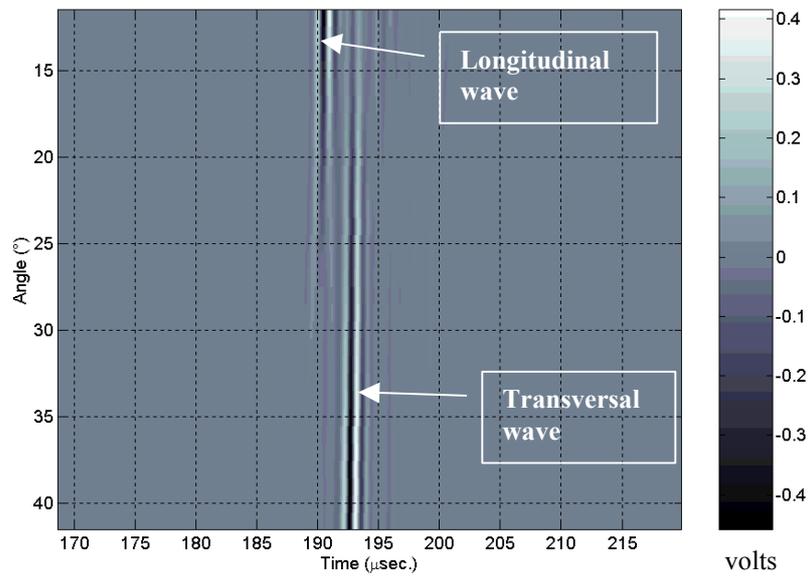

Fig.4

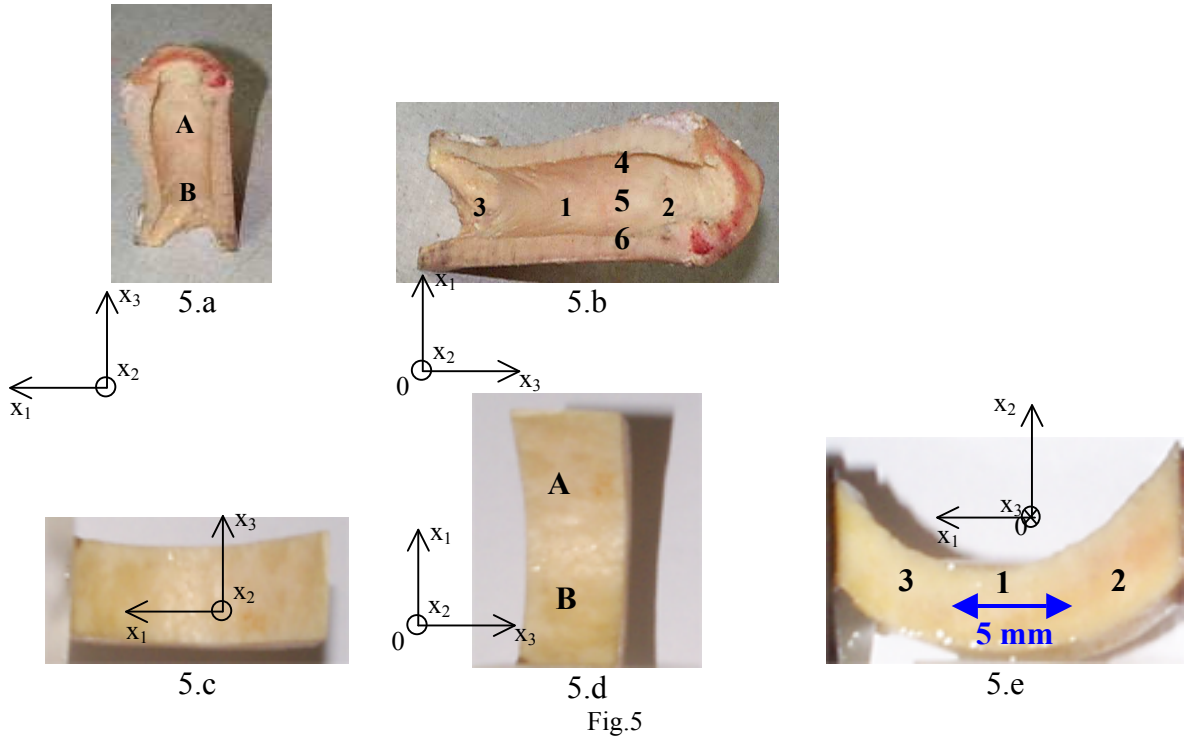

Fig.5

List of tables

Table 1

| Experiments | | $C_L$ (m/s) | $C_T$ (m/s) |
|---|---|---|---|
| Bone 1: (horizontal position) | Point 1 | 3315±20 | 2002±34 |
| | Point 2 | 3655±23 | 1888±28 |
| | Point 3 | 3191±22 | 2026±38 |
| Bone 1: (vertical position) | Point A | 3055±18 | 1767±29 |
| | Point B | 3460±21 | 1773±25 |
| Bone 2: (vertical position) | Point A | 3414±16 | 1822±25 |
| | Point B | 3430±16 | 1833±26 |
| Bone 2: (horizontal position) | Point 1 | 3352±16 | 1953±24 |
| | Point 2 | 3347±16 | 1953±50 |
| | Point 3 | 3317±16 | 1953±41 |
| Bone 2: (horizontal position) | Point 4 | 3487±21 | 2067±44 |
| | Point 5 | 3352±20 | 1949±40 |
| | Point 6 | 3583±21 | 1997±36 |

Table 2

| Experiments | | Transverse isotropic | | Orthotropic | |
|---|---|---|---|---|---|
| | | $C_L$ (m/s) | $C_T$ (m/s) | $C_L$ (m/s) | $C_T$ (m/s) |
| Bone 3 (longitudinal direction, axis 0x) | | 4340±20 | 2062±45 | 4271±20 | 2065±50 |
| Bone 3 (radial direction, vertical position, plane 0xz) | | 3308±15 | 1717±24 | 3235±15 | 1686±20 |
| Bone 3 (radial direction, horizontal position, plane 0xy) | | 3244±15 | 1991±46 | 3321±15 | 1981±31 |
| Bone 3 (radial direction, axis 0y) | | | | 3515±16 | 2082±40 |
| Bone 4 (longitudinal direction, axis 0x) : | Point 1 | 4100±19 | 1976±40 | 4027±13 | 1981±35 |
| | Point 2 | 4326±20 | 2094±58 | | |
| | Point 3 | 4042±19 | 1991±52 | | |
| Bone 4 (radial direction, vertical position, plane 0xz) | | 3388±16 | 1747±27 | 3349±11 | 1730±20 |
| Bone 4 (radial direction, horizontal position, plane 0xy) : | Point A | 3429±16 | 1918±46 | 3350±11 | 1976±37 |
| | Point B | 3350±16 | 1917±46 | | |
| Bone 4 (radial direction (axis 0y)) | | | | 3472±11 | 1973±40 |

Table 3

| | Transverse isotropic | | Orthotropic | |
|---|---|---|---|---|
| | **Bone 3** | **Bone 4** | **Bone 3** | **Bone 4** |
| $C_{11}$(GPa) | 23.3 | 22.4 | 23.5 | 22 |
| $C_{22}$(GPa) | | | 26 | 23.5 |
| $C_{12}$(GPa) | 6.28 | 7.54 | 6.55 | 7.6 |
| $C_{13}$(GPa) | 8.4 | 7.2 | 8.35 | 7.5 |
| $C_{23}$(GPa) | | | 8.2 | 7.7 |
| $C_{33}$(GPa) | 31 | 31 | 34.6 | 31.7 |
| $C_{44}$(GPa) | 9 | 7 | 9.2 | 7.6 |
| $C_{55}$(GPa) | | | 6 | 5.6 |
| $C_{66}$(GPa) | 6.28 | 5.96 | 6.05 | 5.8 |
| $E_1$ (GPa) | 20 | 19 | 20.6 | 18.7 |
| $E_2$ (GPa) | | | 23.4 | 20 |
| $E_3$ (GPa) | 26.2 | 27.5 | 30.2 | 28 |
| $G_{12}$ (GPa) | 3.14 | 3 | 3 | 2.9 |
| $G_{13}$ (GPa) | 4.5 | 3.5 | 3 | 2.8 |
| $G_{23}$ (GPa) | | | 4.6 | 3.7 |
| $\nu_{12}$ | 0.2 | 0.28 | 0.12 | 0.26 |
| $\nu_{13}$ | 0.22 | 0.16 | 0.2 | 0.17 |
| $\nu_{21}$ | | | 0.21 | 0.28 |
| $\nu_{23}$ | | | 0.18 | 0.17 |
| $\nu_{31}$ | | | 0.29 | 0.26 |
| $\nu_{32}$ | | | 0.24 | 0.25 |

Table 4

| | Pithioux, et al. | Katz et al, 1984 | Yoon, et al. 1976 | Ashman, et al., 1984 |
|---|---|---|---|---|
| $C_{11}$(GPa) | $21 \leq C_{11} \leq 24$ | 21.2 (±0.5) | 23.4 (±0.0031) | 18.0 |
| $C_{22}$(GPa) | $23 \leq C_{22} \leq 27$ | 21 (±1.4) | 24.1 (±0.0035) | 20.2 |
| $C_{33}$(GPa) | $28 \leq C_{33} \leq 39$ | 29 (±1) | 32.5 (±0.0044) | 27.6 |
| $C_{44}$(GPa) | $7 \leq C_{44} \leq 9$ | 6.3 (±0.4) | 8.71 (±0.0013) | 6.23 |
| $C_{55}$(GPa) | $5 \leq C_{55} \leq 6$ | 6.3 (±0.2) | 6.9 (±0.0012) | 5.61 |
| $C_{66}$(GPa) | $5 \leq C_{66} \leq 7$ | 5.4 (±0.2) | 7.17 (±0.0011) | 4.52 |
| $C_{12}$(GPa) | $6 \leq C_{12} \leq 11$ | 11.7 (±0.7) | 9.06 (±0.0038) | 9.98 |
| $C_{13}$(GPa) | $7 \leq C_{13} \leq 15$ | 11.1 (±0.8) | 9.11 (±0.0055) | 10.1 |
| $C_{23}$(GPa) | $6 \leq C_{23} \leq 8$ | 12.7 (±0.8) | 9.23 (±0.0055) | 10.7 |